\documentclass[11pt]{article}
\usepackage{fullpage,epsf,amssymb,amsthm,amsfonts,amsmath,latexsym, graphicx,array,extarrows,mathtools,mathstyle,mathrsfs,slashed,subcaption,amsbsy}
\usepackage[normalem]{ulem}
\usepackage{authblk}
\usepackage{cite}
\usepackage{color}

\newcommand{\uvec}[1]{\boldsymbol{#1}}

\title{\bf{Universality of the Poincar\'e gravitational form factor constraints}}

\author[1]{C\'{e}dric Lorc\'{e}\thanks{cedric.lorce@polytechnique.edu}}
\author[1]{Peter Lowdon\thanks{peter.lowdon@polytechnique.edu}}

\affil[1]{\footnotesize{\textit{CPHT, CNRS, Ecole Polytechnique, Institut Polytechnique de Paris, Route de Saclay, 91128 Palaiseau, France}}}

\date{}
\begin{document}

{\let\newpage\relax\maketitle}
\setcounter{page}{1}
\pagestyle{plain}

\abstract

\noindent
Relativistic spin states are convention dependent. In this work we prove that the zero momentum-transfer limits of the leading two form factors in the decomposition of the energy-momentum tensor matrix elements are independent of this choice. In particular, we demonstrate that these constraints are insensitive to whether the corresponding states are massive or not, and that they arise purely due to the Poincar\'e covariance of the states.

\newpage
\section{Introduction}

The form factors that appear in the Lorentz covariant decomposition of the energy-momentum tensor (EMT), the so-called gravitational form factors (GFFs), enter into the physics of many different phenomena, including gravitational scattering~\cite{Boulware:1974sr,Donoghue:2001qc} and the internal properties of hadrons, such as mass, spin and pressure~\cite{Ji:1994av,Ji:1995sv,Ji:1996ek,Polyakov:2002yz,Goeke:2007fp,Leader:2013jra,Roberts:2016vyn,Lorce:2017wkb,Lorce:2017xzd,Polyakov:2018zvc,Lorce:2018egm,Burkert:2018bqq, Kumericki:2019ddg}. In recent years there has been a significant drive to characterise the properties of these objects for target states of different spin~\cite{Polyakov:2002yz,Goeke:2007fp,Lorce:2017wkb,Polyakov:2018zvc,Lorce:2018egm,Kumano:2017lhr,Abidin:2008ku,Taneja:2011sy,Cosyn:2019aio,Polyakov:2019lbq}, due to their connection to generalised parton distributions (GPDs)~\cite{Diehl:2003ny}. By constraining GPDs via the GFFs, this could help, for example, in providing new insights into the dynamics of quarks and gluons within composite states. \\

\noindent
In~\cite{Bakker:2004ib} it was first pointed out that whilst previous studies had correctly identified the constraints imposed on the GFFs in the spin-$\tfrac{1}{2}$ case, these analyses suffered from various technical problems, leading to incorrect physical conclusions. The main issues concerned the treatment of boundary terms, and the non-normalisable momentum eigenstates appearing the EMT matrix elements. These issues were subsequently addressed in~\cite{Lowdon:2017idv} by adopting a non-perturbative distributional approach. This allowed the GFF constraints to be derived in an explicit manner without having to make reference to a particular frame. An important conclusion was that the constraints imposed on the GFFs were not specifically related to linear or angular momentum conservation, but in fact arose due to the Poincar\'e transformation properties of the states. A natural question to ask is whether these constraints continue to hold for massive states of higher spin. Higher spin states have a greater number of potential structures appearing in the EMT matrix element decomposition, as shown for example in~\cite{Cosyn:2019aio} for the spin-1 case. Applying the approach of~\cite{Lowdon:2017idv} on a case-by-case basis would therefore inevitably lead to increasingly more complicated calculations as the spin increased. In~\cite{Cotogno:2019xcl} it was demonstrated that the GFF decomposition can in fact be written in a spin-representation-independent manner, and this ultimately enabled the GFF constraints to be derived for massive states of arbitrary spin. It turned out that the constraints observed in~\cite{Lowdon:2017idv} for the spin-$\tfrac{1}{2}$ case continued to hold for states of \textit{any} spin, and this similarly followed from the Poincar\'e covariance of the states. \\

\noindent
It is well-known that relativistic spin states are convention-dependent. In all previous studies in the literature, including~\cite{Cotogno:2019xcl}, it was implicitly assumed that the states in the EMT matrix elements corresponded to \textit{canonical} spin states, defined by the action of a specific boost on a rest-frame state~\cite{Bakker:2004ib,Polyzou:2012ut}. Although not explicitly stated, the form of this boost transformation played an essential role in deriving the constraints in~\cite{Cotogno:2019xcl}. It therefore remains unclear as to whether the GFF constraints are actually dependent upon the spin-state convention. Moreover, in each of these analyses the states were always assumed to be massive. For massless states several complications are known to arise, including the non-existence of a rest frame, and the necessity to use helicity to define the dynamical degrees of freedom~\cite{Haag:1992hx,Weinberg:1995mt}. As with the spin convention, this naturally leads to the question of whether the GFF constraints are also sensitive to whether the states in the EMT matrix elements are massive or not. The purpose of this work is to address these two questions, and in doing so complete the full classification of constraints imposed on the GFFs. \\

\noindent
The remainder of this paper is structured as follows. In Sec.~\ref{gen_emt} we outline the construction of arbitrary on-shell states, and use this general formulation to define the corresponding EMT matrix element decomposition. Using an analogous approach to that developed in~\cite{Cotogno:2019xcl}, in Sec.~\ref{gen_constr} we derive the general form factor constraints. In Sec.~\ref{concl} we summarise our key findings.

\newpage

\section{Gravitational form factors for arbitrary states}
\label{gen_emt}

The goal of this work is to build on the insights developed in~\cite{Cotogno:2019xcl}, where general constraints were derived on the GFFs for massive canonical states with arbitrary spin. In particular, we will focus on the generalisation of these results to all possible on-shell states, with any mass, spin representation, or spin definition. The foundation of this analysis comes from the fact that any spin state $|p ,\sigma \rangle$ can be defined in the following manner~\cite{Haag:1992hx,Weinberg:1995mt}
\begin{align}
|p , \sigma\rangle = U(L(p))|k,\sigma\rangle,
\label{spin_def}
\end{align}
where $L(p)$ is a choice of Lorentz transformation which maps some reference frame four-vector $k$ to an arbitrary (on-shell) four-momentum $p$
\begin{align}
\Lambda(L(p))k = p,
\end{align}
and thus depends implicitly on the choice of $k$. $U$ is a unitary representation of the Lorentz group, and $\Lambda$ is the corresponding four-vector representation. For massive states, $\sigma$ labels the rest frame spin projection along some axis, whereas for massless states $\sigma$ corresponds to the helicity projection of the state along the direction of motion. Eq.~\eqref{spin_def} emphasises that \textit{any} spin state is uniquely characterised by both the reference vector $k$ and the transformation $L(p)$. Some important examples include~\cite{Bakker:2004ib,Polyzou:2012ut}:
\vspace{3mm}
\begin{itemize}
\item \textit{Canonical spin states} -- states with mass $m$ where $k$ is the rest frame four-vector $(m,0,0,0)$, and $L_{\text{c}}(p)$ is a pure boost along the direction $\hat{\uvec{p}} = \frac{\uvec{p}}{|\uvec{p}|}$.

\item \textit{Wick helicity states} -- massive or massless states where $L_{\text{W}}(p)$ is a pure boost along the $z$-direction followed by a rotation into $\hat{\uvec{p}}$. In the massless case the reference four-vector is defined by $k= (\kappa,0,0,\kappa)$, where $\kappa >0$.

\item \textit{Light-front spin states} -- massive or massless states where $L_{\text{LF}}(p)$ corresponds to a pure boost along the $z$-direction, followed by a transverse light-front boost. 
\end{itemize}
\ \\

\noindent
With these general state definitions one can now define the form factor decomposition of the EMT in an analogous manner to~\cite{Cotogno:2019xcl}. Before doing so, it is important to first consider the general transformation properties satisfied by these states. Firstly, for any state $|p , \sigma\rangle$ of total spin $s$ one can prove from the definition of this state that it transforms under general Lorentz transformations $\alpha$~\cite{Haag:1992hx,Weinberg:1995mt} as:
\begin{align}
U(\alpha)|p, \sigma \rangle = \sum_{\sigma'}\mathcal{D}^{(s)}_{\sigma'\sigma}(W(\alpha,p))|\Lambda(\alpha)p,\sigma' \rangle,
\label{spin_transf}
\end{align} 
where $W(\alpha,p)$ defines a subgroup of reference-frame-preserving Lorentz transformations
\begin{align}
\Lambda(W(\alpha,p))k = k.
\end{align}
This subgroup of \textit{Wigner rotations} $W(\alpha,p)$ is called the little group; for massive states this is $SU(2)$ (the double cover of $SO(3)$) and for massless states this is the Euclidean group $E(2)$. The $p$-dependence of $W(\alpha,p)$ non-trivially depends on the choice of $L(p)$ in the following manner:
\begin{align}
W(\alpha,p) = L^{-1}(\Lambda(\alpha)p) \cdot  \alpha \cdot L(p).
\end{align}
$\mathcal{D}^{(s)}_{\sigma'\sigma}$ is the so-called the \textit{Wigner rotation matrix}, and defines a representation of the little group due to the relation
\begin{align}
U(W(\alpha,p))|k, \sigma \rangle = \sum_{\sigma'}\mathcal{D}^{(s)}_{\sigma'\sigma}(W(\alpha,p))|k ,\sigma' \rangle.
\end{align} 
Since the \textit{generalised polarisation tensors}\footnote{Here we coin the term \textit{generalised polarisation tensor} (GPT) to refer to the Lorentz-index-carrying coefficients $\eta_{\sigma}(p)$ that appear in the canonical free field decomposition. For example, for spin-$\tfrac{1}{2}$ Dirac fields the GPT is simply the Dirac spinor $u_{\sigma}(p)$, and in the spin-1 case the GPT corresponds to the polarisation vector $\varepsilon_{\sigma}(p)$.} (GPTs) $\eta_{\sigma}(p)$ appear in the form factor decomposition of the EMT, it is important to understand how these objects transform under Lorentz transformations. It turns out that in order for general fields $\varphi_{i}(x)$ (with internal index $i$) to transform in a covariant manner
\begin{align}
U(\alpha)\varphi_{i}(x)U^{-1}(\alpha) = \sum_{j}D_{ij}(\alpha^{-1})\varphi_{j}(\Lambda(\alpha)x),
\label{field_rep}
\end{align}  
with $D$ the finite-dimensional Lorentz representation of the field, the GPTs $\eta_{\sigma}(p)$ must satisfy analogous properties to the spin states in Eqs.~\eqref{spin_def} and~\eqref{spin_transf}:
\begin{align}
\eta_{\sigma}(p) &= D(L(p))\eta_{\sigma}(k), \label{eta_rel} \\
D(W)\eta_{\sigma}(k) &= \sum_{\sigma'}\mathcal{D}^{(s)}_{\sigma'\sigma}(W)\eta_{\sigma'}(k),
\end{align}
where the internal indices are omitted here for convenience. In this case though, the action of the Lorentz group on $\eta_{\sigma}(p)$ is determined by the field representation $D$, as opposed to the state representation $U$. By defining the GPTs to have the reference frame normalisation: $\overline{\eta}_{\sigma'}(k)\eta_{\sigma}(k) = \delta_{\sigma'\sigma}$, it immediately follows that    
\begin{align}
\overline{\eta}_{\sigma'}(k)D(W)\eta_{\sigma}(k) = \mathcal{D}^{(s)}_{\sigma'\sigma}(W), 
\label{U_wigner}
\end{align}
which is an important relation for the calculations later in this work. \\

\noindent
Now that the general transformation properties of the states have been outlined, one can proceed to define the corresponding EMT form factor decomposition. For the transformation properties of the states $|p,\sigma\rangle$ to make sense one must implicitly assume that they are on shell. To make this explicit one can define the states by
\begin{align}
|p ,\sigma; M \rangle = \delta_{M}^{(+)}(p)|p,\sigma\rangle \equiv 2\pi \,\theta(p^{0})\,\delta(p^{2}-M^{2}) |p,\sigma\rangle,
\end{align}   
where either $M>0$ or $M=0$, and $|p,\sigma\rangle$ has unrestricted momentum $p$. All of the previously derived relations in this section continue to hold with these states. For our purposes we are interested in the symmetric EMT matrix element. It turns out that due to the conservation of $T^{\mu\nu}$, together with the assumption that the current is both hermitian and invariant under discrete ($\mathsf{P}$ and $\mathsf{T}$) symmetries, the matrix element of the EMT has the decomposition~\cite{Boulware:1974sr,Cotogno:2019xcl}:   
\begin{align}
\langle p',\sigma';M|T^{\mu\nu}(0)|p ,\sigma;M \rangle = \overline{\eta}_{\sigma'}(p')O^{\mu\nu}(p',p)\eta_{\sigma}(p)  \, \delta_{M}^{(+)}(p')\,\delta_{M}^{(+)}(p),
\label{T_decomp}
\end{align}
where $O^{\mu\nu}(p',p)$ is a Lorentz covariant operator which acts on the GPTs and has the following representation-independent form:
\begin{align}
O^{\mu\nu}(p',p)= \bar{p}^{\{\mu}\bar{p}^{\nu\}}  A(q^{2}) + i \bar{p}^{\{\mu}\widetilde{D}(S^{\nu\}\rho})q_{\rho} \, G(q^{2}) + \cdots
\label{O_decomp}
\end{align}  
The $\cdots$ represent contributions with an explicitly higher-order dependence on the four-momentum transfer $q=p'-p$. The index symmetrisation is defined: $a^{\{\mu }b^{\nu\}}= a^{\mu}b^{\nu}+a^{\nu}b^{\mu}$, and $\bar{p}= \tfrac{1}{2}(p'+p)$. EMT conservation demands that each term in Eq.~\eqref{O_decomp} vanishes when contracted with $q$, hermiticity requires invariance under complex conjugation and $q \rightarrow -q$, and the discrete symmetries imply that the second term must involve both $\bar{p}$ and $q$. \\  

\noindent
$S^{\mu\nu}$ in Eq.~\eqref{O_decomp} are the \textit{abstract} Lorentz group generators, and $\widetilde{D}$ is the corresponding Lie algebra representation of $D$ introduced in Eq.~\eqref{field_rep}, defined via the exponential map: $D\left(\text{exp}(X)\right) = \text{exp}\left(\widetilde{D}(X)\right)$ for any Lorentz generator $X$. As will be demonstrated in the remainder of this work, the subtle differences between the various Lorentz group and algebra representations play an important role in the analysis of the GFF constraints. In the next section we will perform an analogous procedure to~\cite{Cotogno:2019xcl}, using the matrix elements of both the non-covariant and covariant Lorentz generators to derive constraints on $A(q^{2})$ and $G(q^{2})$, but this time using the state-independent quantities defined in this section.

\section{Lorentz generator matrix elements for arbitrary states}
\label{gen_constr}

\subsection{Angular momentum matrix element}

Following the same logic as in~\cite{Cotogno:2019xcl}, the decomposition in Eq.~\eqref{T_decomp} implies that the angular momentum matrix element has the form\footnote{See~\cite{Cotogno:2019xcl} for more details about the specific calculation, including the definition of the rotation generator $J^{i}$.}  
\begin{align}
\langle p',\sigma';M|\widetilde{U}(J^{i})|p ,\sigma ;M\rangle = (2\pi)^{4}\delta_{M}^{(+)}(\bar{p})\, \mathcal{J}^{i}_{\sigma'\sigma}(\bar{p},q),
\end{align}
with the reduced matrix element
\begin{align}
\mathcal{J}^{i}_{\sigma'\sigma}(\bar{p},q) &= -i\epsilon^{ijk}\bar{p}^{k} \left[ \delta_{\sigma'\sigma} \, \partial^{j}\delta^4(q) - \partial^{j} \!\left[\overline{\eta}_{\sigma'}(p')\eta_{\sigma}(p)\right]_{q=0} \delta^4(q)\right]  A(q^2) \nonumber \\
&\quad + \left[ \overline{\eta}_{\sigma'}(\bar{p})\widetilde{D}(J^{i})\eta_{\sigma}(\bar{p}) \right] \delta^4(q) \,    G(q^2).
\label{red_J}
\end{align}
Here we make explicit that the abstract rotation generator $J^{i}$ acts on the states via the Lie algebra representation $\widetilde{U}$ associated with $U$, and on the GPTs via the representation $\widetilde{D}$. Using Eq.~\eqref{eta_rel} one can now write the reduced matrix element in a manner that explicitly depends on the choice of $L(p)$. Firstly, one has that
\begin{align}
\overline{\eta}_{\sigma'}(p')\eta_{\sigma}(p) &= \overline{\eta}_{\sigma'}(k)D^{-1}(L(p'))D(L(p))\eta_{\sigma}(k) \nonumber \\
&= \overline{\eta}_{\sigma'}(k)D\left(L^{-1}(p')L(p) \right)\eta_{\sigma}(k) \nonumber \\
&= \overline{\eta}_{\sigma'}(k)D \left(L^{-1}(\bar{p} + \tfrac{1}{2}q)L(\bar{p} - \tfrac{1}{2}q) \right)\eta_{\sigma}(k), 
\end{align}
where we have used the fact that $D$ is a Lie group homomorphism and switched to the coordinates $(\bar{p},q)$. In order to further simplify Eq.~\eqref{red_J} one needs to evaluate the following expression:
\begin{align}
\left.\frac{\partial}{\partial q_{j}}\right|_{q=0} D \left(L^{-1}(\bar{p} + \tfrac{1}{2}q)L(\bar{p} - \tfrac{1}{2}q) \right).
\label{der_U}
\end{align}
To do so one can make use of the general representation theory identity\footnote{This relation follows from the fact that $\widetilde{\phi}$ is the differential of the map $\phi$ at the identity~\cite{Warner83}.}:
\begin{align}
\left.\frac{d}{dt}\right|_{t=0} \! \phi \left(f(t) \right) = \widetilde{\phi} \left( \left.\frac{df}{dt}\right|_{t=0} \right),
\label{general_ident}
\end{align}
where $f$ is a Lie group-valued function with $f(0)=1$, and $\phi$ is a Lie group representation which (uniquely) defines a corresponding Lie algebra representation $\widetilde{\phi}$. Since $L^{-1}(\bar{p})L(\bar{p} ) = 1$, one can apply the multi-dimensional generalisation of Eq.~\eqref{general_ident} to Eq.~\eqref{der_U}. Doing so, one obtains
\begin{align}
\left.\frac{\partial}{\partial q_{j}}\right|_{q=0} \!D \left(L^{-1}(\bar{p} + \tfrac{1}{2}q)L(\bar{p} - \tfrac{1}{2}q) \right) &= \widetilde{D} \! \left( \frac{1}{2}\frac{\partial L^{-1}(\bar{p})}{\partial \bar{p}_{j}} L(\bar{p}) - \frac{1}{2}L^{-1}(\bar{p})\frac{\partial L(\bar{p})}{\partial \bar{p}_{j}} \right) \nonumber \\
&= \widetilde{D} \! \left( \frac{\partial L^{-1}(\bar{p})}{\partial \bar{p}_{j}} L(\bar{p}) \right),
\end{align} 
where the final equality follows from the fact that: $\frac{\partial L^{-1}(\bar{p})}{\partial \bar{p}_{j}} L(\bar{p}) = - L^{-1}(\bar{p})\frac{\partial L(\bar{p})}{\partial \bar{p}_{j}}$. By also making use of the representation theory identity: $D(L^{-1})\widetilde{D}(J^{i})D(L) = \widetilde{D}\left(L^{-1}J^{i}L\right)$, the reduced matrix element can finally be written 
\begin{align}
\mathcal{J}^{i}_{\sigma'\sigma}(\bar{p},q) = &-i\epsilon^{ijk}\bar{p}^{k} \delta_{\sigma'\sigma} \, \partial^{j}\delta^{4}(q) A(q^2) +i\epsilon^{ijk}\bar{p}^{k} \,  \overline{\eta}_{\sigma'}(k)\widetilde{D} \! \left( \frac{\partial L^{-1}(\bar{p})}{\partial \bar{p}_{j}} L(\bar{p}) \right)\eta_{\sigma}(k)  \,  \delta^{4}(q)  A(q^2) \nonumber \\
& + \overline{\eta}_{\sigma'}(k)\widetilde{D}\left(L^{-1}(\bar{p})J^{i}L(\bar{p})\right)\eta_{\sigma}(k) \, \delta^{4}(q) \,  G(q^2),
\label{red_J_2}
\end{align}
where the dependence on the state definition $L$ is now explicit. \\

\noindent
To derive constraints on the form factors one can use the transformation properties of the states under a pure rotation about the $i$ axis ($\alpha= \mathcal{R}_{i}= e^{-i \beta J^{i}}$) to write an alternative representation of the reduced matrix element $\mathcal{J}^{i}_{\sigma'\sigma}$. In this notation, one finds that
\begin{align}
\mathcal{J}^{i}_{\sigma'\sigma}(\bar{p},q) = -i\epsilon^{ijk}\bar{p}^{k} \delta_{\sigma'\sigma} \, \partial^{j}\delta^4(q) + i \,\delta^{4}(q) \left.\frac{d}{d \beta}\right|_{\beta=0} \! \mathcal{D}^{(s)}_{\sigma'\sigma}(W(\mathcal{R}_{i},\bar{p})),
\end{align}
where the two terms arise due to the Lorentz transformation dependence of both the Wigner rotation matrix and the state in Eq.~\eqref{spin_transf}. To compare this with Eq.~\eqref{red_J_2} one needs to rewrite this in terms of the GPTs $\eta_{\sigma}(p)$. For this purpose one can apply Eq.~\eqref{U_wigner}, which gives 
\begin{align}
\mathcal{J}^{i}_{\sigma'\sigma}(\bar{p},q) =  -i\epsilon^{ijk}\bar{p}^{k} \delta_{\sigma'\sigma} \, \partial^{j}\delta^4(q) + i \,\delta^{4}(q) \left.\frac{d}{d \beta}\right|_{\beta=0} \!  \overline{\eta}_{\sigma'}(k)D(W(\mathcal{R}_{i},\bar{p}))\eta_{\sigma}(k).
\label{red_J_3}
\end{align} 
In this case the corresponding Wigner rotation has the form 
\begin{align}
W(\mathcal{R}_{i}, \bar{p}) = L^{-1}\left(\Lambda\left(\mathcal{R}_{i}\right) \bar{p}\right) e^{-i \beta J^{i}} L(\bar{p}). 
\end{align}
Since $W(\beta=0)=1$, one can similarly apply Eq.~\eqref{general_ident} in order to simplify the second term in Eq.~\eqref{red_J_3}. Applying the chain rule one finds that 
\begin{align}
i \!\left.\frac{d}{d \beta}\right|_{\beta=0} \! W(\mathcal{R}_{i}, \bar{p}) = \left[\widetilde{\Lambda}( J^{i})\bar{p} \right]^{\!\mu}\frac{\partial L^{-1}(\bar{p})}{\partial \bar{p}^{\mu}}L(\bar{p}) + L^{-1}(\bar{p})J^{i}L(\bar{p}),
\end{align}
where $\widetilde{\Lambda}$ is the Lie algebra representation associated with the four-vector representation $\Lambda$. Since $\left[ \widetilde{\Lambda}(J^{i})\bar{p} \right]^{\!\mu} = -ig^{\mu}_{\ l}\epsilon^{ilk}\bar{p}^{k}$ and $\widetilde{D}$ is a linear map, one can write
\begin{align}
i \!\left.\frac{d}{d \beta}\right|_{\beta=0} \! D \!\left( W(\mathcal{R}_{i}, \bar{p}) \right) &= \widetilde{D} \!\left( i\epsilon^{ijk}\bar{p}^{k}\frac{\partial L^{-1}(\bar{p})}{\partial \bar{p}_{j}}L(\bar{p}) + L^{-1}(\bar{p})J^{i}L(\bar{p}) \right) \nonumber \\
&=  i\epsilon^{ijk}\bar{p}^{k} \, \widetilde{D} \!\left( \frac{\partial L^{-1}(\bar{p})}{\partial \bar{p}_{j}}L(\bar{p}) \right) + \widetilde{D} \!\left( L^{-1}(\bar{p})J^{i}L(\bar{p}) \right). 
\label{U_rot_der}
\end{align}
Finally, combining Eqs.~\eqref{red_J_3} and~\eqref{U_rot_der} one obtains   
\begin{align}
\mathcal{J}^{i}_{\sigma'\sigma}(\bar{p},q) = &-i\epsilon^{ijk}\bar{p}^{k} \delta_{\sigma'\sigma} \, \partial^{j}\delta^{4}(q)  +i\epsilon^{ijk}\bar{p}^{k} \,  \overline{\eta}_{\sigma'}(k)\widetilde{D} \! \left( \frac{\partial L^{-1}(\bar{p})}{\partial \bar{p}_{j}} L(\bar{p}) \right)\eta_{\sigma}(k)  \,  \delta^{4}(q)  \nonumber \\
& + \overline{\eta}_{\sigma'}(k)\widetilde{D}\left(L^{-1}(\bar{p})J^{i}L(\bar{p})\right)\eta_{\sigma}(k) \, \delta^{4}(q). 
\label{red_J_4}
\end{align}
Comparing this representation with Eq.~\eqref{red_J_2} it immediately follows that the form factors must satisfy the following constraints
\begin{align}
A(q^{2})\,\delta^{4}(q) &= \delta^{4}(q), \label{constr1} \\
A(q^{2})\,\partial^{j}\delta^{4}(q) &= \partial^{j}\delta^{4}(q), \label{constr2} \\
G(q^{2})\,\delta^{4}(q) &= \delta^{4}(q), \label{constr3}
\end{align}
which are identical to those derived in~\cite{Cotogno:2019xcl} for massive canonical states. By using the most general representation of on-shell relativistic spin states we have proven that $A(0)$ and $G(0)$ are \textit{completely independent} of the characteristics of the states which define them. This result emphasises that the Poincar\'{e} transformation properties of the states alone are sufficient to constrain that: $A(0)=G(0)=1$. Among other things, this implies that Ji's sum rule~\cite{Ji:1996ek} is completely independent of the states used in the definition of the corresponding GPDs, and that the vanishing of the anomalous gravitomagnetic moment~\cite{Teryaev:1999su} is not only true for arbitrary spin, but is also a mass-independent condition. In other words, these results demonstrate that the implications for massive canonical states derived in~\cite{Cotogno:2019xcl} are actually a specific realisation of the fact that the GFF constraints are \textit{state universal}. In contrast to~\cite{Cotogno:2019xcl}, these constraints were derived without ever having to make reference to the explicit form of the transformation $L(p)$. This circumvents the significant complications required in taking the derivative of $L(p)$, as well as having to determine the angular momentum matrix element in different spin conventions, which in the massless Wick helicity case is known to be particularly complicated~\cite{Bakker:2004ib}.

\subsection{Boost matrix element}

Similarly to the angular momentum matrix element, in the boost case one can write
\begin{align}
\langle p',\sigma';M|\widetilde{U}(K^{i})|p ,\sigma ;M\rangle = (2\pi)^{4}\delta_{M}^{(+)}(\bar{p})\, \mathcal{K}^{i}_{\sigma'\sigma}(\bar{p},q),
\end{align}
with the reduced matrix element\footnote{In~\cite{Cotogno:2019xcl} the term involving the temporal derivative of the GPT product in Eq.~\eqref{boost_red} was not written because it explicitly vanishes for massive canonical spin states ($L=L_{\text{c}}$). However, for general spin-states $L(p)$ this term need not vanish.} given by
\begin{align}
\mathcal{K}^{i}_{\sigma'\sigma}(\bar{p},q) &= i \left[\delta_{\sigma'\sigma}\,(\bar{p}^{0} \partial^{i} - \bar{p}^{i} \partial^{0})\delta^4(q) - (\bar{p}^{0} \partial^{i} - \bar{p}^{i} \partial^{0})\!\left[\overline{\eta}_{\sigma'}(p')\eta_{\sigma}(p)\right]_{q=0}\delta^{4}(q)       \right]  A(q^{2}) \nonumber \\
& \quad +\left[\overline{\eta}_{\sigma'}(\bar{p})\widetilde{D}(K^{i})\eta_{\sigma}(\bar{p})  \right] \delta^{4}(q)  \, G(q^{2}).
\label{boost_red}
\end{align} 
Analogously to the angular momentum case this can be rewritten as
\begin{align}
&\mathcal{K}^{i}_{\sigma'\sigma}(\bar{p},q) = i \delta_{\sigma'\sigma}\,(\bar{p}^{0} \partial^{i} - \bar{p}^{i} \partial^{0})\delta^4(q) \, A(q^{2}) - i \bar{p}^{0} \, \overline{\eta}_{\sigma'}(k)\widetilde{D} \! \left( \frac{\partial L^{-1}(\bar{p})}{\partial \bar{p}_{i}}L(\bar{p}) \right)\eta_{\sigma}(k) \,  \delta^4(q) \, A(q^{2})  \nonumber \\
& +i \bar{p}^{i} \, \overline{\eta}_{\sigma'}(k)\widetilde{D} \! \left( \frac{\partial L^{-1}(\bar{p})}{\partial \bar{p}_{0}} L(\bar{p}) \right)\eta_{\sigma}(k) \,  \delta^{4}(q) \, A(q^{2})  +\overline{\eta}_{\sigma'}(k)\widetilde{D}\left(L^{-1}(\bar{p})K^{i}L(\bar{p})\right)\eta_{\sigma}(k) \,  \delta^{4}(q)  \, G(q^{2}).
\label{boost_red_2}
\end{align} 
This time one must instead use the transformation properties of the states under a pure boost along the $i$-direction ($\alpha= \mathcal{B}_{i}= e^{i \xi K^{i}}$) to write the reduced matrix element in an alternative representation. After applying Eq.~\eqref{U_wigner} one finds that
\begin{align}
\mathcal{K}^{i}_{\sigma'\sigma}(\bar{p},q) = i\delta_{\sigma'\sigma}\, (\bar{p}^{0} \partial^{i} - \bar{p}^{i} \partial^{0})\delta^4(q) -i \,\delta^{4}(q) \left.\frac{d}{d \xi}\right|_{\xi=0} \!  \overline{\eta}_{\sigma'}(k)D(W(\mathcal{B}_{i},\bar{p}))\eta_{\sigma}(k). 
\label{boost_red_3}
\end{align}
The next step is to calculate the derivative of the Wigner rotation for $\mathcal{B}_{i}$, which gives 
\begin{align}
-i \!\left.\frac{d}{d \xi}\right|_{\xi=0} \! W(\mathcal{B}_{i}, \bar{p}) = \left[\widetilde{\Lambda}( K^{i})\bar{p} \right]^{\!\mu}\frac{\partial L^{-1}(\bar{p})}{\partial \bar{p}^{\mu}}L(\bar{p}) + L^{-1}(\bar{p})K^{i}L(\bar{p}).
\end{align}
By using the relation $\left[ \widetilde{\Lambda}(K^{i})\bar{p} \right]^{\!\mu} = ig^{\mu 0}\bar{p}^{i}  -ig^{\mu i}\bar{p}^{0}$ together with Eq.~\eqref{general_ident} and the linearity of $\widetilde{D}$, the reduced matrix element can finally be written in the form
\begin{align}
&\mathcal{K}^{i}_{\sigma'\sigma}(\bar{p},q) = i \delta_{\sigma'\sigma}\,(\bar{p}^{0} \partial^{i} - \bar{p}^{i} \partial^{0})\delta^{4}(q)  - i \bar{p}^{0}\, \overline{\eta}_{\sigma'}(k)\widetilde{D} \! \left( \frac{\partial L^{-1}(\bar{p})}{\partial \bar{p}_{i}} L(\bar{p}) \right)\eta_{\sigma}(k) \,  \delta^{4}(q)  \nonumber \\
& +i \bar{p}^{i} \, \overline{\eta}_{\sigma'}(k)\widetilde{D} \! \left( \frac{\partial L^{-1}(\bar{p})}{\partial \bar{p}_{0}} L(\bar{p}) \right)\eta_{\sigma}(k) \,  \delta^{4}(q) \, A(q^{2})  +\overline{\eta}_{\sigma'}(k)\widetilde{D}\left(L^{-1}(\bar{p})K^{i}L(\bar{p})\right)\eta_{\sigma}(k) \,  \delta^{4}(q). 
\label{boost_red_4}
\end{align} 
Comparing Eqs.~\eqref{boost_red_2} and~\eqref{boost_red_4} one is then immediately led to identical constraints to those in Eqs.~\eqref{constr1}--\eqref{constr3}, together with the condition: $A(q^{2})\partial^{0}\delta^{4}(q) = \partial^{0}\delta^{4}(q)$. 

\subsection{Pauli-Lubanski matrix element}

As opposed to $J^{i}$ and $K^{i}$, the operator form of the Pauli-Lubanski operator $W^{\mu}$ depends on whether the states are massive or massless. Using the form factor decomposition for \textit{massive} states, the reduced matrix element of $W^{\mu}$ is given by 
\begin{align}
\mathcal{W}^{\mu}_{\sigma'\sigma}(\bar{p},q) = \overline{\eta}_{\sigma'}(k)\widetilde{D}\left(L^{-1}(\bar{p})W^{\mu}L(\bar{p})\right)\eta_{\sigma}(k) \, \delta^{4}(q) \, G(q^{2}), \label{W_red_1}
\end{align}
where\footnote{Here we use the convention $\epsilon_{0123}=+1$.} $W^{\mu}= \frac{1}{2}\epsilon^{\mu}_{\phantom{\mu}\rho\sigma\lambda}S^{\rho\sigma}\bar{p}^{\lambda}$. Since $W^{\mu}$ depends on both the angular momentum and boost operators, one can use the state-independent relations in Eqs.~\eqref{red_J_4} and~\eqref{boost_red_4} to derive a general representation for the matrix elements of $W^{\mu}$. For $\mu=0$
\begin{align}
\mathcal{W}^{0}_{\sigma'\sigma}(\bar{p},q) &= \bar{p}^{i} \mathcal{J}^{i}_{\sigma'\sigma}(\bar{p},q) \nonumber \\
&= \bar{p}^{i} \left[ -i\epsilon^{ijk}\bar{p}^{k} \delta_{\sigma'\sigma} \, \partial^{j}\delta^{4}(q)  +i\epsilon^{ijk}\bar{p}^{k} \,  \overline{\eta}_{\sigma'}(k)\widetilde{D} \! \left( \frac{\partial L^{-1}(\bar{p})}{\partial \bar{p}_{j}} L(\bar{p}) \right)\eta_{\sigma}(k) \,  \delta^{4}(q) \right] \nonumber \\
& \quad + \bar{p}^{i} \, \overline{\eta}_{\sigma'}(k)\widetilde{D}\left(L^{-1}(\bar{p})J^{i}L(\bar{p})\right)\eta_{\sigma}(k) \, \delta^{4}(q) \nonumber \\
&=  \overline{\eta}_{\sigma'}(k)\widetilde{D}\left(L^{-1}(\bar{p})\, \bar{p}^{i}J^{i}L(\bar{p})\right)\eta_{\sigma}(k) \, \delta^{4}(q),
\label{W_red_0}
\end{align}
where the first term vanishes due to the contraction with $\epsilon^{ijk}\bar{p}^{k}$. For $\mu=i$ one finds
\begin{align}
\mathcal{W}^{i}_{\sigma'\sigma}(\bar{p},q) &=  \bar{p}^{0}\mathcal{J}^{i}_{\sigma'\sigma}(\bar{p},q) + \epsilon^{ijk}\bar{p}^{k} \mathcal{K}^{j}_{\sigma'\sigma}(\bar{p},q) \nonumber \\
&= \bar{p}^{0}\left[ -i\epsilon^{ijk}\bar{p}^{k} \delta_{\sigma'\sigma} \, \partial^{j}\delta^{4}(q)  +i\epsilon^{ijk}\bar{p}^{k} \,  \overline{\eta}_{\sigma'}(k)\widetilde{D} \! \left( \frac{\partial L^{-1}(\bar{p})}{\partial \bar{p}_{j}} L(\bar{p}) \right)\eta_{\sigma}(k) \,  \delta^{4}(q) \right] \nonumber \\
& \quad +\overline{\eta}_{\sigma'}(k)\widetilde{D}\left(L^{-1}(\bar{p})\, \bar{p}^{0}J^{i}L(\bar{p})\right)\eta_{\sigma}(k) \, \delta^{4}(q) \nonumber \\
& \quad + \epsilon^{ijk}\bar{p}^{k} \bigg[i \delta_{\sigma'\sigma}\,(\bar{p}^{0} \partial^{j} - \bar{p}^{j} \partial^{0})\delta^{4}(q)  - i \bar{p}^{0}\, \overline{\eta}_{\sigma'}(k)\widetilde{D} \! \left( \frac{\partial L^{-1}(\bar{p})}{\partial \bar{p}_{j}} L(\bar{p}) \right)\eta_{\sigma}(k) \,  \delta^{4}(q)  \nonumber \\
&\quad +i \bar{p}^{j} \, \overline{\eta}_{\sigma'}(k)\widetilde{D} \! \left( \frac{\partial L^{-1}(\bar{p})}{\partial \bar{p}_{0}} L(\bar{p}) \right)\eta_{\sigma}(k) \,  \delta^{4}(q)  \bigg]  \nonumber \\
& \quad +  \overline{\eta}_{\sigma'}(k)\widetilde{D}\left(L^{-1}(\bar{p})\epsilon^{ijk}\bar{p}^{k}K^{j}L(\bar{p})\right)\eta_{\sigma}(k) \,  \delta^{4}(q) \nonumber \\
&= \overline{\eta}_{\sigma'}(k)\widetilde{D}\left(L^{-1}(\bar{p})\left[\bar{p}^{0}J^{i} +\epsilon^{ijk}\bar{p}^{k}K^{j} \right]L(\bar{p})\right)\eta_{\sigma}(k) \,  \delta^4(q).
\label{W_red_i}
\end{align}
Taken together, Eqs.~\eqref{W_red_0} and~\eqref{W_red_i} therefore imply the general representation
\begin{align}
\mathcal{W}^{\mu}_{\sigma'\sigma}(\bar{p},q) =  \overline{\eta}_{\sigma'}(k) \, \widetilde{D}\! \left(L^{-1}(\bar{p})W^{\mu}L(\bar{p})\right)\eta_{\sigma}(k) \, \delta^4(q).
\label{W_red_2}
\end{align}
Comparing this with Eq.~\eqref{W_red_1} one is then led to the constraint
\begin{align}
G(q^{2})\,\delta^4(q) = \delta^4(q).
\label{G_Contr}
\end{align}
This calculation generalises the findings in~\cite{Cotogno:2019xcl}, proving that the covariantised operator $W^{\mu}$ constraining only $G(q^{2})$ is not just a characteristic of massive canonical spin states, but in fact occurs for any choice of massive spin state. \\

\noindent
For \textit{massless} states one has instead that the Pauli-Lubanski operator is proportional to the energy-momentum vector
\begin{align}
W^{\mu} = H P^{\mu} = \frac{\uvec{P}\cdot \uvec{J}}{|\uvec{P}|}P^{\mu}
\end{align}
where $H$ is the helicity operator. Inserting this between massless states, and comparing it with the corresponding form factor expression, one obtains Eq.~\eqref{G_Contr}, analogously to the $W^{0}$ component in the massive case.    

\subsection{Covariant boost matrix element}

To complete the analysis in this section we consider the matrix elements of the covariant boost operator $B^{\mu}$, defined by
\begin{align}
B^{\mu} = \frac{1}{2}\left[ S^{\nu\mu}P_{\nu} + P_{\nu}S^{\nu\mu} \right].
\end{align}
It follows from this definition that the temporal and spatial components of the reduced matrix elements have the form 
\begin{align}
\mathcal{B}^{0}_{\sigma'\sigma}(\bar{p},q) &= \bar{p}^{i} \, \mathcal{K}^{i}_{\sigma'\sigma}(\bar{p},q), \label{reducedB1} \\
\mathcal{B}^{i}_{\sigma'\sigma}(\bar{p},q) &= \bar{p}^{0} \, \mathcal{K}^{i}_{\sigma'\sigma}(\bar{p},q) +\epsilon^{ijk}\bar{p}^{j} \, \mathcal{J}^{k}_{\sigma'\sigma}(\bar{p},q).
\label{reducedB2}
\end{align}
After substituting in the form factor expressions for the reduced matrix elements in Eqs.~\eqref{red_J_2} and~\eqref{boost_red_2} one finds that the coefficient of $G(q^{2})$ has the following general form
\begin{align}
\overline{\eta}_{\sigma'}(k) \, \widetilde{D}\! \left(L^{-1}(\bar{p})B^{\mu}L(\bar{p})\right)\eta_{\sigma}(k). 
\label{G_coeff}
\end{align}
For \textit{massive} states the covariant boost $B^{\mu}$ is related to the rest frame boost $B^{\mu}_{k} \equiv (0,K^{i})$ by
\begin{align}
B^{\mu}_{k} = L^{-1}(\bar{p})B^{\mu}L(\bar{p}).
\end{align}
The coefficient of $G(q^{2})$ therefore reduces to: $\overline{\eta}_{\sigma'}(k) \, \widetilde{D}\! \left(B^{\mu}(k)\right)\eta_{\sigma}(k)$, which vanishes\footnote{This occurs because massive physical states have a vanishing intrinsic energy dipole moment~\cite{Lorce:2018zpf}.}, and hence the covariant boost matrix element only results in constraints on $A(q^{2})$. Taken together with the results for $W^{\mu}$, this proves that the diagonalisation of constraints observed for canonical spin states in~\cite{Cotogno:2019xcl} is actually independent of the (massive) spin state convention. For \textit{massless} states though, the coefficient in Eq.~\eqref{G_coeff} will generally not vanish, and so an analogous diagonalisation does not occur.

\section{Conclusions}
\label{concl}

Although the constraints imposed on the gravitational form factors have been explored for massive canonical spin states of lower spin, and more recently for arbitrary spin~\cite{Cotogno:2019xcl}, it remained an open question as to whether these constraints depended on the choice of spin state definition, or if the states were massive or not. In this work we definitively answer this question, proving that the zero-momentum transfer constraint $A(0)=G(0)=1$ of the leading two form factors is independent of both of these conditions, and arises purely due to the Poincar\'e covariance of the states. Besides the relevance for hadronic physics, the universality of this constraint could also potentially have important implications for the understanding of gravitational scattering amplitudes. 

\section*{Acknowledgements}

This work was supported by the Agence Nationale de la Recherche under the Projects No. ANR-18-ERC1-0002 and No. ANR-16-CE31-0019.

\bibliographystyle{JHEP}

\bibliography{paper_refs}

\end{document}